\documentclass[prl,twocolumn,showpacs,amsmath,amssymb]{revtex4}

\def\e{{\,\rm e}\,}
\def\d{{\rm d}}
\def\K{{K}}
\def\i{{\rm i}}
\newcommand{\rf}[1]{(\ref{#1})}
\newcommand{\eq}[1]{Eq.~(\ref{#1})}
\def\be{\begin{equation}}
\def\ee{\end{equation}}
\def\bea{\begin{eqnarray}}
\def\eea{\end{eqnarray}}
\def\LA{\left\langle}
\def\RA{\right\rangle}
\def\t{{s}}
\def\D{{\cal D}}
\def\Tau{{\cal T}}
\newcommand{\non}{\nonumber \\*}

\def\la{\lesssim}
\def\ga{\gtrsim} 

\begin{document}

\title{Implementation of the Duality between Wilson loops  \\
and Scattering Amplitudes in QCD}

\author{Yuri Makeenko}
\altaffiliation[Also at]{ the Institute for Advanced Cycling,
Blegdamsvej 19, 2100 Copenhagen \O, Denmark}

\affiliation{Institute of Theoretical and Experimental Physics,
Moscow, Russia}

\author{Poul Olesen$^*$} 
\affiliation{The Niels Bohr International Academy,
The Niels Bohr Institute,
Copenhagen, Denmark}


\begin{abstract}
We generalize  modern ideas about the duality between Wilson loops and 
scattering amplitudes in ${\cal N}=4$ SYM to large-$N$ (or quenched) QCD.
We show that the area-law behavior of asymptotically large Wilson
loops is dual to the Regge--Veneziano behavior of scattering amplitudes
at high energies and fixed momentum transfer, when quark mass is small 
and/or the number of particles is large.
We elaborate on this duality for string theory in flat space,
identifying the asymptotes of the disk amplitude and the Wilson loop of 
large-$N$ QCD.

\end{abstract}


\maketitle


This Letter is inspired by a remarkable recent discovery of the duality
between Wilson loops and scattering amplitudes in ${\cal N}=4$ super 
Yang--Mills (SYM) theory (see~\cite{AR08} for a review of this subject).
SYM differs from QCD by the contents of matter fields (6 scalars and 4 spinors 
in the adjoint representation of the SU($N$) color group, thereby providing
an extended  ${\cal N}=4$ supersymmetry) and has attracted a significant 
interest over the last three decades as a toy model for certain
aspects of QCD, in particular, for the relation between QCD and strings.
Adding the extra fields makes the dynamics of SYM much simpler than that
of QCD and it enjoys the famous 
anti--de-Sitter-space/conformal-field-theory 
(AdS/CFT) correspondence which, in particular, relates 
SYM Wilson loops to an
open superstring in AdS$_5\otimes S^5$~\cite{Mal98}. 
While essential ingredients of QCD --- asymptotic freedom and quark 
confinement --- are not present in ${\cal N}=4$ SYM, 
it captures many features of QCD 
perturbation theory.

The (finite part of the) 4-gluon on-shell scattering
amplitude in SYM has the form 
\be
A(s,t)=A_{\rm tree}\,\e^{f(\lambda) \log^2 (s/t)}
\label{BDS}
\ee
(where $s$ and $t$ are usual Mandelstam's variables)  
as was conjectured~\cite{BDS05} on the basis
of three-loop calculations.
To explain \eq{BDS}, the 
Wilson-loop/scattering-amplitude (WL/SA) duality 
was introduced~\cite{AM07a} at large 't~Hooft
couplings $\lambda$, which has been then advocated in SYM perturbation
theory~\cite{DSK07}. This duality states that the scattering
amplitude (divided by the kinematical factor $A_{\rm tree}$) equals
the Wilson loop for a rectangle whose vertices $x_i$ are related to
the momenta $p_i$ of scattering gluons by
\be
p_i= \K\left( x_i-x_{i-1}\right),
\label{du}
\ee
where $\K=1/2\pi\alpha^\prime$ is the string tension.

The function $f(\lambda)$ 
also appears in the anomalous dimensions of cusped Wilson loops and 
operators of twist two. It 
has been recently 
found as a solution to 
the equation~\cite{ES06} derived from spin chains. 
Its perturbative solution reproduces the three~\cite{KLOV,BDS05}
and four~\cite{BCDKS} loop SYM results, while numerical~\cite{BBKS} and
analytical~\cite{BKK07} solutions reproduce the $\sqrt{\lambda}$ behavior
of $f(\lambda)$ for large $\lambda$ originally found~\cite{GKP02} 
using the AdS/CFT correspondence.  
The next orders in $1/\sqrt{\lambda}$ also agree with 
the superstring calculations~\cite{FT02}
thereby providing a remarkable test of the AdS/CFT correspondence. 

Our goal in this Letter is to find out what features of the described
WL/SA duality (if any) remain valid for QCD and, in particular,
how is it possible to maintain the relation of the type~\rf{du}
which would relate {\em large}\/ momenta in scattering amplitudes with
loops of {\em large}\/ size. Of course this is not possible in QCD
perturbation theory, 
where $|p|\sim 1/|x|$ because of
dimensional ground. But nonperturbatively a dimensional parameter 
$\K\approx (400~{\rm MeV})^2$
appears in QCD, which shows up
in the area-law behavior of asymptotically large Wilson loops:
\be
W(C)\stackrel{{\rm large}~C} \propto \e^{-\K S_{\rm min}(C)} \,,
\label{a-l}
\ee
where $S_{\rm min}(C)$ is the area of the minimal surface bounded by $C$,
that results in confinement. Strictly speaking, this requires 
large 
$N$ or the quenched approximation.

As is well-known by now, 
a string theory, which QCD is supposedly equivalent
to, is not the simplest Nambu--Goto string. Some
extra degrees of freedom living on the string are required
which are most probably
conveniently described by a presence of extra dimensions. The asymptotic
behavior~\rf{a-l} is nevertheless universal for large loops. Also,
there is a considerable amount of evidence from
lattice gauge calculations in 2+1 and 3+1 dimensions for various $N$
that the Nambu--Goto action describes the behavior of the Wilson loops
quite well 
and the transition
from perturbative to stringy behavior takes place ``at surprisingly
small distances''~\cite{lw02}. There also exists a number of
other comparisons between results from the Nambu--Goto action, e.g. between 
the closed string spectrum, and SU($N$) for various $N$
(see \cite{close} and references therein).
This
action has the well-known anomaly for $d\neq 26$, which however is suppressed
for long strings~\cite{p85}. 

This remarkable success of the Nambu--Goto string in flat space
as an effective action leads us to reconsider the relation between the Wilson
loop $W(C)$ and the corresponding string wave functional.
We then obtain scattering amplitudes in large-$N$ (or quenched) QCD by
properly summing $W(C)$ over paths and find that the WL/SA duality holds
in a kinematical region of large $s$ and fixed $t$ when only
{\em large}\/ loops, for which the area law~\rf{a-l} sets in, are essential
in the sum over paths. Thus obtained scattering amplitudes,
quite involved in general, 
are of the Regge--Veneziano type when the quark mass is 
small and/or the number of external particles is large.


Our starting point is the standard representation of
Green's functions of $M$ colorless composite quark
operators (e.g. ${\bar q}(x_i) q (x_i)$) in terms of the sum
over all Wilson loops passing via the points $x_i$ ($i=1,\ldots,M$), where the
operators are inserted:
\begin{eqnarray}
&&\hspace*{-5mm}
G\equiv \LA \prod_{i=1}^M \bar{q} (x_i) q (x_i) \RA_{\rm conn} =
\int\nolimits_0^\infty \d \Tau\, \e^{-m \Tau} \non &&\hspace*{-4mm}\times
\int\nolimits_0^{\Tau} \d \tau_{M-1}
\prod_{i=1}^{M\!-\!2}\int\nolimits_0^{\tau_{i\!+\!1}} \!\!\d \tau_i \hspace*{-6mm}
\int\limits_{z(0)=z(\Tau)=x_{0} \atop z(\tau_i)=x_{i}} \hspace*{-6mm}\D z(\tau) \,
J[z(\tau)] \, W[z(\tau)].~~~
\label{115}
\end{eqnarray}
Here the weight for the path integration is
\be
J[z(\tau)]=\int \D k(\tau)\;{\rm sp~P} 
\e^{\i \int_0^{\Tau} \d \tau\,
[\dot z (\tau)\cdot k(\tau)-\gamma(\tau)\cdot k(\tau)]} 
\label{J}
\ee
for spinor quarks and scalar operators.
In \eq{115} $W(C)$ is the Wilson loop in pure Yang--Mills theory at large $N$
(or quenched), $m$ is the quark mass and $\tau$ is the proper-time variable.
For finite $N$, correlators of several Wilson loops have to be taken into 
account.
The derivation of this formula and the references can be found 
in~\cite{Mak02}.

The on-shell $M$-particle scattering amplitudes can be obtained from the 
Green function~\rf{115} by the standard Lehman--Symanzik--Zimmerman 
reduction.
When making the Fourier transformation, it is convenient to
represent $M$ momenta of the (all incoming) particles by the
differences
$ %
\Delta p_i= p_{i\!-\!1}-p_i\,.
$ 
Then momentum conservation is automatic while an (infinite) volume 
$V$ is produced, say, by integration over $x_0$. 
It is convenient to introduce a momentum-space loop
$p_\mu(\tau)$ which is piecewise constant:
\begin{equation}
p(\tau)=p_i \qquad \hbox{for}~ \tau_i<\tau<\tau_{i\!+\!1} \,.
\label{piecewise}
\end{equation}
Because    
 $\dot { p}(\tau) =- \sum_i \Delta  p_i \delta(\tau-\tau_i)$ with 
$\Delta p_i \equiv  p_{i\!-\!1} -  p_i$,
we write 
in the Fourier transformation:
$ 
\sum_i \Delta p_i \cdot x_i = 
 \int \d \tau \,p(\tau) \cdot \dot z(\tau)
$ 
which is manifestly parametric invariant.

Making the Fourier transformation, we obtain
\begin{eqnarray}
&&\hspace*{-4mm}
G\left(\Delta p_1,\ldots, \Delta p_M \right)=
\int\nolimits_0^\infty \d \Tau\, \e^{-m \Tau} 
\int\nolimits_0^{\Tau} \d \tau_{M-1}\non&&\hspace*{-4mm}\times \!
\prod_{i=1}^{M\!-\!2}\int\nolimits_0^{\tau_{i\!+\!1}} \!\!\d \tau_i  \hspace*{-6mm}
\int\limits_{z(0)=z(\Tau)=0} \hspace*{-6mm} \D z(\tau) \,
\e^{\i \int_0^\Tau \d \tau\, \dot z (\tau)\cdot p(\tau)}\,
J[z(\tau)] \, W[z(\tau)],  \nonumber \\[-5mm] && \mbox{}
\label{117}
\end{eqnarray}
where $p(\tau)$ is piecewise constant as in \eq{piecewise}. 
We do not integrate over $z(0)=z(\Tau)$ which would produce 
the (infinite) volume factor because of translational invariance. 


To calculate the scattering amplitudes, we have to substitute the
area-law behavior~\rf{a-l} of asymptotically large Wilson loops into \eq{117}
and to integrate over the paths. In general, this would lead us to
very complicated integrals but the calculation drastically simplifies
if to use the representation of the minimal area as a boundary
functional that was introduced by Douglas~\cite{Dou31} in his
celebrated solution of the Plateau problem.
We shall use one of the equivalent forms of the Douglas functional:
\bea
A[\sigma]&=&
-\frac 1{4\pi} \int_0^{\Tau}\d\tau_1 \d\tau_2\,
\dot x(\tau_1)\cdot \dot x(\tau_2)\, \non && \times 
\ln \left(1-\cos\left[{2\pi}
\left(\sigma(\tau_1)-\sigma( \tau_2)\right)/\Tau\right]  \right),~~
\label{D'}
\eea
where $0<\sigma(\tau)<\Tau$ is a reparametrization ($\sigma^\prime(\tau)\geq 0$).
The functional~\rf{D'} is to be minimized with respect to $\sigma(\tau)$
with the minimizing function 
$\sigma_*(\tau)$ being of course contour-dependent. 
Then $A[\sigma_*]$ is equal to the minimal area $S_{\rm min}(C)$,
while in general $A[\sigma]\geq A[\sigma_*]=S_{\rm min}(C)$.  

In fact~\rf{D'} is well-known as the classical boundary action in string theory.
It appears for the tree-level disk amplitude with Dirichlet boundary
conditions in the Polyakov string formulation 
after integrating over the string fluctuations inside the disk,
i.e.\ over $ X(r,\theta)$ with $r< 1$, $0\leq\theta< 2\pi$, 
and fixing the value $ X(1,\theta) \equiv  x(\theta)$
at the boundary. The appearance of the function $\sigma_*(\theta)$
is related to a subtlety associated with fixing conformal 
gauge~\cite{Alv83}. The 
decoupling of the Liouville field 
is possible only in the interior of
the disk, while its boundary value determines the
function $\sigma_*(\theta)$ at the classical level.
The path integral over the boundary value of the Liouville field then
restores the invariance under reparametrizations of the boundary in quantum
theory.


Motivated by this fact, Polyakov~\cite{Pol97} proposed 
to identify the  Wilson loop in large-$N$ QCD
with the tree-level string disk amplitude integrated over
reparametrizations of the boundary contour.
It is convenient to conformally  map
the disk into the upper half-plane, so the disk boundary 
is mapped into the real axis parametrized by 
$t(\tau)= \tan (\pi \tau/\Tau)$,
$-\infty <t < +\infty$.
Then we write
\bea
W(C)&=&  
\int \D \t(t) \exp{\Big(\,\frac \K{2\pi}
\int\nolimits_{-\infty}^{+\infty} {\d t_1 \d t_2} \, 
\dot x(t_1) \cdot \dot x(t_2)} \non &&~~~{\times
\ln |\t(t_1)-\t(t_2)|   \Big)},
\label{diskx}
\eea
where the path integral over $\t(t)$ (with $\t^\prime(t)\geq 0$) 
restores the invariance under reparametrizations.

In spite of the fact that the right-hand side of \eq{diskx} is derivable
for bosonic string in $d=26$ or superstring in $d=10$,
we shall use it only for asymptotically large loops or, equivalently,
very large $\K$, when the integral over
reparametrizations has a saddle point at $s(t)=s_*(t)$.
This is crucial for reproducing \eq{a-l}.


It is easy to calculate a (reparametrization-invariant) functional 
Fourier transformation 
\be
W[ p(\cdot)]= \int \D  x\, \e^{\i \int  p \cdot\d x }\;
W [x(\cdot)]
\label{Fourier}
\ee
of the disk amplitude~\rf{diskx}
for piecewise constant 
$p(t)$.
Substituting~\rf{diskx} into \eq{Fourier} and performing the Gaussian
integration, we get 
\bea
W[p(\cdot)]&=&  
\int \D s(t) \exp{\Big({\alpha^\prime}
\int\nolimits_{-\infty}^{+\infty} {\d t_1 \d t_2} \, 
\dot p(t_1) \cdot \dot p(t_2)} \non &&~~~{\times
\ln |s(t_1)-s(t_2)|   \Big)}
\label{diskp}
\eea
which is 
of the same form as \rf{diskx} only with 
$\K$ replaced by $1/\K=2\pi\alpha^\prime$.

Since $ p(t)= p_j$ at the $j$-th interval for the stepwise
discretization, the only effect of the
reparametrization is to change the values of $t_j$'s for $\t_j$'s 
keeping their cyclic order.
This is a discrete version of the reparametrization transformation.
Note that the stepwise discretization of $ x(t)$ itself is not possible
since it would violate the continuity of the world-line of the string end. 

The stepwise discretization~\rf{piecewise} naturally  results
in the $M$-particle (off-shell) Koba--Nielsen amplitudes 
which are invariant under the
SL$(2;\mathbb{R})$ projective transformation 
$\t\Rightarrow ({a \t+b})/({c \t+d})$ with  $a d-b c=1 $
because the projective group is a subgroup of reparametrization
transformations.
To derive them, we first note that
\bea
&&\hspace*{-4mm}\mbox{}\int\nolimits_{-\infty}^{+\infty} {\d t_1 \d t_2} \, 
\dot p(t_1) \cdot \dot p(t_2) \ln |s(t_1)-s(t_2)| \non
&&=
-\frac 12
\int\nolimits_{-\infty}^{+\infty} \frac{\d \t_1 \d \t_2}{(\t_1-\t_2)^2} \, 
\left[p\left(t(s_1)\right)- p\left(t(s_2)\right)\right]^2~~~
\eea
for the integral in the exponent in \rf{diskp}.
The integration over $\t_1$ or $\t_2$  on the right-hand side has divergences 
when they lie on
adjacent sides $k=l\pm 1$. If we omit the sides
with $k=l\pm1$, then  the integrations over $\t_1$ and $\t_2$ are
perfectly finite resulting in
\bea
\lefteqn{\hspace*{-5mm}\frac 12
\sum_{k\neq l\pm1}\; \int\nolimits_{\t_{k-1}}^{\t_k} \d \t_1
\int\nolimits_{\t_{l-1}}^{\t_l} \d \t_2\, \frac{(p_k-p_l)^2}{(\t_1-\t_2)^2}} 
 \nonumber \\&=& 
\sum_{k\neq l\pm1}\; \Delta p_k \cdot\Delta p_l \log |\t_k-\t_l| \non &&
+ \sum_j \Delta p_j^2 \log 
\frac{(\t_j-\t_{j-1})(\t_{j+1}-\t_j)}{(\t_{j+1}-\t_{j-1})}
\eea
which is projective invariant.


Choosing the measure to be 
\be
\D \t=\prod_i \d\t_i\frac{(\t_{i+1}-\t_{i-1})}{(\t_i-\t_{i-1})(\t_{i+1}-\t_i)}
\label{msr}
\ee
which is also invariant under the projective transformation,
we arrive at
\begin{eqnarray}
&&\hspace*{-3mm}W(\Delta p_1,\ldots,\Delta p_M)=\hspace*{-3mm}
\int\limits_{\t_{j-1}<\t_j} \prod_i \d \t_i
\prod_{k\neq l} |\t_k-\t_l|^{\alpha^\prime\Delta  p_k\Delta  p_l }\non &&~~~~~ 
~~~\times
\prod_j\left(\frac{(\t_j-\t_{j-1})(\t_{j+1}-\t_j)}{(\t_{j+1}-\t_{j-1})}\right)
^{\alpha^\prime \Delta p_j^2-1},
\label{Lovelace}
\end{eqnarray}
where the integration over $\t_j$ emerges from the path integral 
over reparametrizations in \eq{diskp}.
This is known as the Lovelace choice~\cite{Lov70} (see~\cite{DiV92}),
that reproduces some projective-invariant
off-shell string amplitudes known since late 1960's.
The more familiar on-shell tachyon amplitudes can be obtained
from \eq{Lovelace} by setting $\alpha^\prime \Delta  p_j^{2}=1$.  

Fixing in \eq{Lovelace} the remaining SL$(2;\mathbb{R})$ invariance 
in the standard way,
we obtain the scalar amplitudes in the Koba--Nielsen variables.
For the  case of 4 scalars this reproduces the Veneziano amplitude
\be
A(\Delta p_1,\Delta p_2,\Delta p_3,\Delta p_4)
=\int\nolimits_0^1 \d x\, x^{-\alpha(s)-1}
(1-x)^{-\alpha( t)-1} ,
\label{veneziano}
\ee
\vspace*{-2mm} \mbox{}\\ where $\alpha(s)=\alpha^\prime s+1$ and
$
s=-(\Delta p_1+\Delta p_2)^2$, 
\mbox{$t=-(\Delta p_2+\Delta p_3)^2$}
are usual Mandelstam's variables (for Euclidean metric).
Here the tachyonic condition $\alpha^\prime \Delta p_j^2=1$
has not to be imposed. 
While \eq{Lovelace} results in $\alpha(0)=1$,
an arbitrary value of the intercept $\alpha(0)$ can be reached by 
properly changing the measure~\rf{msr}.


We are now in a position to perform the main task of this Letter:
to substitute the area-law behavior~\rf{a-l} 
of $W(C)$ into the path integral~\rf{117} and to find out
for what momenta the asymptotically large loops dominate.
As we shall see, typical momenta will be large for large loops.
As is already explained, we substitute \eq{a-l} by \eq{diskx} which
gives the same for large loops (or large $\K= 1/2\pi\alpha^\prime$).

Interchanging the order of integration over $z(\tau)$ and $\sigma(\tau)$ 
(or $s(t)$) 
and easily doing a Gaussian path integral, we obtain
\begin{widetext}
\begin{eqnarray}
G\left(\Delta p_1,\ldots, \Delta p_M \right)&=&  
\int\nolimits_0^\infty \d \Tau\,\e^{-m \Tau} 
\int\nolimits_0^{\Tau} \d \tau_{M-1}
\prod_{i=1}^{M\!-\!2}\int\nolimits_0^{\tau_{i\!+\!1}} \d \tau_i  
\,\int \D \sigma (\tau) \int\D k(\tau) \nonumber \\ 
&& ~\times {\rm sp~P}\,
\e^{\frac {\alpha^\prime}{2}\int_0^{\Tau} \d\tau_1 \int_0^{\Tau} \d\tau_2 
(\dot{k}(\tau_1)+\dot{p}(\tau_1))\cdot  (\dot{k}(\tau_2)+\dot {p}(\tau_2)) 
\ln\left(1-\cos\left[\frac{2\pi}{\Tau}\left(\sigma(\tau_1)-\sigma(\tau_2)\right)
\right]\right)
-\i\int_0^{\Tau}\d\tau\, \gamma (\tau)\cdot k(\tau)}.~~~~
\label{sves1}
\end{eqnarray}
\end{widetext}
This expression is rather close 
to the disk amplitude~\rf{diskp}, 
except for the additional integration over $k$. 
But for the case where $m$ 
is small (or $M$ is very large), the integral over $\Tau$ in \eq{sves1} 
is dominated by $\Tau\sim (M-1)/m$ which is large for $m$ small. 
This is because $\prod_{i}^{M-1} \d \tau_i \sim \Tau^{M-1}$.
Noting that typical
values of $k\sim 1/\Tau$ are essential in the path integral over $k$
for large $\Tau$, we can disregard $k(\tau)$ in
the exponent in \eq{sves1} so the integral over $k$ factorizes.
Making the change of the variables from $\sigma$ to $\t$, 
we finally obtain from \eq{sves1} the product of the momentum-space
disk amplitude~\rf{diskp} times factors which do not depend on $p$.
Therefore, \eq{sves1} exactly reproduces for piecewise constant $p(\tau)$
the (off-shell) Koba--Nielsen amplitude~\rf{Lovelace} as 
$m\rightarrow0$!

It still remains to discuss in what kinematical region of momenta $\Delta p_i$
our derivation is legible, that is only
asymptotically large loops are essential in the path
integral over $z(\tau)$ in \eq{117}. 
A physical intuition suggests, from the spectrum of
a classical string, this should be the case at least for 
asymptotically large $s$ and large $t\la s$.
This indeed agrees with our formulas, where the value of $\alpha(0)$
is not essential in this region. But the domain of applicability of
our approach is broader and extends to large negative values of $t$.
However, when $-t \ll s$ becomes large enough, 
there are no longer reasons to expect the contribution of large loops to 
dominate over perturbation theory, which comes from integration over
small loops in \eq{117}. Therefore, our formula 
for the 4-point scattering amplitude is valid only 
for asymptotically large $s$ and fixed $t$ ($|t|\ga 1/\alpha^\prime$), 
associated with {\it small}\/ angle or fixed momentum transfer.
The tachyon issue, which is  
a short distance phenomenon~\cite{p85}, is then irrelevant.
For smaller values of $|t|\la 1/\alpha^\prime$ the results become sensitive to
the choice of the measure in the ansatz~\rf{diskx}.

As distinct from previous approaches to reggeization in perturbative QCD,
in particular from that based on the evolution equation~\cite{KK} for 
Regge trajectories, 
our approach deals with large loops 
usually associated with nonperturbative effects.
Actually we are dealing with the quark-antiquark Regge trajectory,
whose QCD calculation was pioneered in~\cite{KL83}, rather
than with the Pomeron. 

When $m$ is not small and/or $M$ is not large, one should consider the full 
expression (\ref{sves1}). 
We can split there the 
$k-$integral into two domains with small and large $\dot{k}$. Then the
former will appear as a Regge--Veneziano behaved factor coupled to the 
rest of the integrand. 

Thus, in conclusion we see that 
{\it the area-law behavior of Wilson loops is dual 
to the Regge--Veneziano behavior of scattering amplitudes 
at high energies and small angles, when quark mass is small 
and/or the number of produced particles is large}, 
but this ceases to be valid when the momentum transfer is large.
This is how the exponential falloff of the 
4-particle amplitude with large
$-t\sim s$, which is unavoidable in string theory~\cite{GM87}, does
not happen in our consideration. 

{\it Acknowledgments.}---We thank R.~Marotta for pointing 
out that the amplitude~\rf{Lovelace} is projective invariant.

\end{document}